\documentstyle[12pt]{article}

\def\p{{\rm {\bf p}}}
\def\q{{\rm {\bf q}}}

\begin{document}

\hfill BI-TP 96/13

\hfill March 1996

\vspace{1.5cm}

\begin{center}
{\bf HOT QUARK-GLUON MATTER WITH DECONFINED HEAVY QUARKS }
\footnote{Research is partially supported by "Volkswagen-Stiftung"}

\vspace{1.5cm}
{\bf O.K.Kalashnikov}
\footnote{Permanent address: Department of
Theoretical Physics, P.N.Lebedev Physical Institute, Russian
Academy of Sciences, 117924 Moscow, Russia. E-mail address:
kalash@td.lpi.ac.ru}

Fakult\"at f\"ur Physik

Universit\"at Bielefeld

D-33501 Bielefeld, Germany

\vspace{2.5cm}

{\bf Abstract}

\end{center}

The phase diagram of the quark-gluon matter evolution is presented 
for the SU(3)-model with a new phase of heavy deconfined quarks which 
exists in a rather wide range of temperatures and densities. Fitting 
the chiral phase transition data to fix the model parameters we 
establish another (deconfinement) phase transition which separates 
a new phase from hadronic matter. The parameters and properties 
of the phase diagram are discussed in comparison with lattice 
and other results.

\newpage

At present for many applications which are based on hot QCD predictions 
and for future experiments it is very essential to establish a reliable 
$(\mu,T)$-phase diagram of the QCD-thermodynamical evolution and to 
calculate quantitatively the parameters which determine its phase 
properties. However this is not the case today since the properties 
of many (at least, two) phases are not elaborated in detail.  Only the
quark-gluon plasma phase (QGP-phase) which takes place at high
temperatures and densities is reliably studied since due to a small
coupling constant it can be treated perturbatively.  This is the phase 
of the "light" and deconfined quarks and gluons which are  
massless in the first approximation. We know as well that there is a 
phase which presents hadronic matter (H-phase) where quarks and gluons 
are confined to form the observable nuclear matter. However, there is 
a question about the intermediate phase (Q-phase) which should present 
heavy and deconfined quarks. This phase is very natural and has a good 
physical meaning but till now it is not established firmly. Of course, 
it has been studied in many papers (in [1] and then in other papers) 
but no reliable predictions have been given. Today, the interest in 
this problem was revived again by paper [2] where the intermediate 
Q-phase is explicitly found and its parameters are determined within 
the bag model ideology.

The goal of this paper is to derive the QCD-pattern of the 
Q-phase directly from the Lagrangian approach using the standard
calculations within the temperature Green function technique. Fitting
the chiral phase transition data to fix the model parameters we 
establish another (deconfinement) phase transition which separates 
the Q-phase from hadronic matter. Here we consider that the quark mass 
sharply arises after crossing the chiral phase transition line and the 
Q-phase demonstrates itself as a phase of heavy quarks with a rather 
strong interaction. The deconfinement phase transition occurs when               
the temperature decreases and within our scenario it 
does not coincide with the chiral transition. There are two 
well-separated phase transitions and below we specify their parameters.

The QCD Lagrangian is usual and in covariant gauges it
has the form
\setcounter{equation}{0}
\begin{eqnarray}
{\cal L}=&-&\frac{1}{4}{G_{\mu\nu}^a}^2+N_f{\bar \psi}
[\gamma_{\mu}(\partial_{\mu}-\frac{1}{2}ig\lambda^aV_{\mu}^a)+m]\psi
\nonumber\\ &-&\mu N_f{\bar \psi}\gamma_4\psi
+\frac{1}{2\alpha}(\partial_{\mu}V_{\mu}^a)^2 +{\bar C}^a
(\partial_{\mu}\delta^{ab}+gf^{abc}V_{\mu}^c)\partial_{\mu}C^b
\end{eqnarray}
where $G_{\mu\nu}^a=\partial_{\mu}V_{\nu}^a-\partial_{\nu}V_{\mu}^a
+gf^{abc}V_{\mu}^bV_{\nu}^c$  is the Yang-Mills field strength;
$V_{\mu}$ is a nonabelian gauge field; $\psi$(and ${\bar
 \psi}$) are the quark fields in the SU(N)-fundamental representation
($\frac{1}{2}\lambda^a $ are its generators and $f^{abc}$ are the
SU(N)-structure constants) and $C^a$ (and ${\bar C}^a$) are the ghost
Fermi fields. In Eq.(1) $\mu$ and $m$ are the quark chemical potential 
and bare quark mass, respectively, $N_f$ is the number of quarks 
flavours and $\alpha$ is the gauge fixing parameter ($\alpha=1$ for 
the Feynman gauge). The metric is chosen to be Euclidean and 
$\gamma_{\mu}^2=1$.

The partition function is determined by the functional integral
in the standard manner [3]
\begin{eqnarray}
Z=N\int {\cal D}(\psi,C,V){\displaystyle\exp}
\left[\int d^3x \int\limits _0^{\beta}dx_4{\cal L}(\psi,C,V)\right]
\end{eqnarray}
where $\beta$ is the inverse temperature and we calculate the
thermodynamical potential for hot QCD following the formula
\begin{eqnarray}
\Omega\;=\;-\;T\;{\displaystyle\ln} Z\;.  
\end{eqnarray}
The two-loop approximation for $\Omega$ is considered at first but
then we go beyond the perturbative theory. In the two-loop approximation 
the four diagrams (or three ones if the axial gauge is used) should be 
calculated. The calculations are performed with the aid of the bare 
Green and vertex functions in the Matsubara technique. The Feynman 
gauge is more preferable although the final result
does not depend on the gauge choice. The renormalization prescription is
well-known [4,5] and due to the quark loop (where is the mass 
renormalization) it is not trivial and more complicated then for hot 
gluodynamics. Such calculations were made many years ago in [4] for hot 
QED where the thermodynamical potential with nonperturbative 
corrections was found and then in [5,6,7] for hot QCD. In a more 
general case, when the quark mass and chemical potential $m,\mu \ne 0$, 
the two-loop thermodynamical potential was calculated in [7] and this 
expression is given by
\begin{eqnarray}
\frac{\Omega^{(2)}}{V}=&-&\frac{\pi^2(N^2-1)}{45}T^4
+\frac{g^2(N^2-1)N}{144}T^4
-\frac{NN_f}{3\pi^2}\int\limits_0^{\infty}\frac{d\p}{E_{\p}}\p^4n_{\p}
\nonumber\\&+&\frac{N_f(N^2-1)g^2T^2}
{24\pi^2} \int\limits_0^{\infty}\frac{d\p}{E_\p}\p^2 n_{\p}
+\frac{N_f(N^2-1)g^2}{32\pi^4}\int\limits_0^{\infty}\frac{d\p d\q}
{E_{\p}E_{\q}}\p^2 \q^2 \nonumber\\
&\times&\left[ \left( 2+\frac{m^2}{\p\q}
{\displaystyle\ln}\frac{E_{\p}E_{\q}-m^2-\p\q}{E_{\p}E_{\q}-m^2+\p\q}
\right)(n_{\p}^-n_{\q}^-+n_{\p}^+n_{\q}^+)\nonumber\right.\\
&+&\left.(n_{\p}^-n_{\q}^+ +n_{\p}^+n_{\q}^-)
\left(2+\frac{m^2}{\p\q}
{\displaystyle\ln}\frac{E_{\p}E_{\q}+m^2+\p\q}{E_{\p}E_{\q}+m^2-\p\q}
\right)\right]
\end{eqnarray}
where $n_{\p}=n_{\p}^+ +n_{\p}^-$  and $n_{\p}^{\pm}=
[{\displaystyle\exp}\beta(E_{\p} \pm \mu ) + 1]^{-1}$ are the usual
quark occupation numbers. Here $E_{\p}=(\p^2+m^2)^{1/2}$ is the quark
energy.

Our goal is to go beyond the perturbative expansion. There are 
several ways of doing it but phenomenologically (as a convenient fit)
it is possible to replace $g^2$ by the running constant $g^2(Q^2)$ 
which further can be used for any temperatures and densities. Here we 
put our fit into agreement with the chiral phase transition data and 
then the Q-phase with the massive quarks will be investigated to find 
the deconfinement phase transition. The proposed fit has a rather 
simple form
\begin{eqnarray}
g^2(Q^2)=\frac{\pi^2b_1}{b_0^2\;{\displaystyle\ln}\Bigl[1+Q^{
{\displaystyle 2(\frac{b_1}{16b_0})}}\Bigr]}
\end{eqnarray}
where $b_0=(11N-2N_f)/3$ and $b_1=(34N^2-13NN_f+3N_f/N)/3$ are the
standard renormalization group coefficients (for other more complicated
fit see  [8]). The ansatz (5) correctly reproduces the 
high momentum (temperature or density) behaviour and demonstrates the
interaction strengthening when a small momentum region is considered.
In this region where the infrared divergencies are dominant
the canonical dimensions are removed from the theory (like in QED see
e.g. in [3])  and anomalous ones arise.

At first, we study the simplest case within the SU(3)-model when 
$\mu,m=0$ to check the fit (5) and to fix the parameter $\Lambda$. 
When the temperature decreases the chiral phase transition should 
occur and namely this case is more convenient for lattice simulations 
and for analytical calculations as well: here all integrals in Eq.(4) 
are calculated exactly. The two-loop thermodynamical potential is  
given by
\begin{eqnarray}
\frac{\Omega^{(2)}}{V}=-\frac{\pi^2T^4}{45}[(N^2-1)+\frac{7}{4}
NN_f]+ g^2T^4\frac{(N^2-1)}{144}(N+\frac{5}{4}N_f)
\end{eqnarray}
however Eq.(6) being a pure perturbative expansion does not demonstrate
any phase transition [9]. The useful model arises only after replacing 
$g^2$ with the aid of Eq.(5) and we study its properties below. 
The phase transition is determined by the condition 
$(-p=\Omega(Q_c)=0)$ which here results in the equation
\begin{eqnarray}
\frac{b_1}{b_0^2\;{\displaystyle\ln}\Bigl[1+Q_c^{
{\displaystyle 2(\frac{b_1}{16b_0})}}\Bigr]} =
\frac{144 [(N^2-1)+\frac{7}{4}
NN_f]}{45(N^2-1)(N+\frac{5}{4}N_f)}               
\end{eqnarray}                     
where $Q_c=T_c/\Lambda$ is a scaled variable and $N=3$. 
Solving this equation one finds the critical $Q_c$-points for 
each $N_f$ as follows (see Table 1).  
\begin{table}[h]           
\centering
\begin{tabular}{|r||r|r|r|}
\hline
$SU(3)$ & $N_f=2$ & $N_f=3$ & $N_f=4$ \\ 
\hline        
$Q_c^2$ & 0.704 & 0.529 & 0.346 \\
$T_c(MeV)$ &  186  & 160   & 130   \\
\hline
\end{tabular}
\caption{The data summary for the chiral phase transition
within the SU(3)-model. Here $\Lambda=\;$222MeV to reproduce 
$T_c=\;$260MeV since $Q_c^2$=1.375 for $N_f=0$. For comparison, 
see lattice data in [10] and about  $\Lambda$ in [11].} 
\end{table}

\noindent
Of course, the data in Table 1 can be improved but the order of their 
values and the tendency (the monotonic decreasing $T_c$ with $N_f$) 
are correct. The proposed model reliably reproduces the chiral 
transition temperature for any $N_f$.

{\it Now the model is fixed } and we study 
the Q-phase with massive quarks to establish the deconfinement phase 
transition when the temperature becomes lower. These calculations are 
more complicated since all integrals in Eq.(4) are not treated exactly 
when the quark mass $m$ is non-zero. However for $\mu=0$  Eq.(4) keeps 
a rather convenient form
\begin{eqnarray}
\frac{\Omega^{(2)}}{V}=&-&\frac{\pi^2(N^2-1)}{45}T^4
-\frac{2NN_fT^4\omega^4}{3\pi^2}I_1(\omega)\nonumber\\
&+&\frac{g^2(N^2-1)N}{144}T^4
+\frac{N_f(N^2-1)g^2T^4}{12}\left[\;\frac{\omega^2}{\pi^2}
I_2(\omega)\nonumber\right.\\
&+&\left.\frac{\omega^4}{\pi^4}\left(\;3\;I_2^2(\omega)
+\frac{3}{4}I_3(\omega)\right)\right]
\end{eqnarray}
since all integrals in this case depend on the scaled parameter 
$\omega=m/T$. These integrals are 
\begin{eqnarray}
I_1(\omega)
&=&\int\limits_0^{\infty}\frac{x^4dx}{\sqrt{x^2+1}}\frac{1}
{{\displaystyle\exp}(\omega\sqrt{x^2+1})+1}\nonumber\\
I_2(\omega)
&=&\int\limits_0^{\infty}\frac{x^2dx}{\sqrt{x^2+1}}\frac{1}
{{\displaystyle\exp}(\omega\sqrt{x^2+1})+1}\nonumber\\
I_3(\omega)
&=&\int\limits_0^{\infty}\frac{xdx}{\sqrt{x^2+1}}
\frac{ydy}{\sqrt{y^2+1}}{\displaystyle\ln}
\frac{(x-y)^2}{(x+y)^2}\nonumber\\&\times&\frac{1}
{{\displaystyle\exp}(\omega\sqrt{x^2+1})+1}
\;\frac{1}{{\displaystyle\exp}(\omega\sqrt{y^2+1})+1}
\end{eqnarray}
and they have to be calculated numerically. But, there is a problem 
with the $I_3$-integral which is not prepared properly for numerical
calculations. However this integral being rather small for 
$\omega\geq 1$ does not affect essentially the final result and due to 
this fact its logarithm can be slightly redefined  
\begin{eqnarray}
{\displaystyle\ln}\frac{(x-y)^2}{(x+y)^2}\rightarrow
{\displaystyle\ln}\frac{(x-y)^2+0.01}{(x+y)^2+0.01}
\end{eqnarray}
to make this integral convenient for the numerical calculations as 
well.

Let us find the critical point $T_d$ which
determines the deconfinement phase transition in our scenario. 
We consider the SU(3)-model with $N_f$=(2 and 3): the   
experimental data seem to be between these values. It is also 
assumed that all constituent quarks have the same masses which are
generated due to the chiral phase transition and have the magnitude
of 180MeV $\leq m\leq $350MeV. This quantity being a free parameter is 
fitted with the aid of the $\omega$-variable to define another 
(deconfinement) phase transition. The equation which 
determines the critical temperature of this transition is established 
through the "standard" requirement ($-p=\Omega(Q_d)=0)$ and using 
Eq.(8) its explicit form is given by
\begin{eqnarray}
&&\frac{8}{45}+\frac{2N_f\omega_d^4}{\pi^4}I_1(\omega_d)
=\frac{g^2(Q_d^2)}{\pi^2}\nonumber\\&&\times\left
\{\frac{1}{6}+\frac{2N_f}{3}\Bigl(\;\frac{\omega_d^2}{\pi^2}\;
I_2(\omega_d)+\frac{\omega_d^4}{\pi^4}\;
[\;3\;I_2^2(\omega_d)+\frac{3}{4}I_3(\omega_d)\;]\;\Bigr)\right\}
\end{eqnarray}
where $\omega_d =m/T_d$ and $Q_c=T_d/\Lambda$. In Eq.(11) all 
integrals are  calculated numerically for the chosen 
$\omega_d$-values and then the equation obtained is solved to 
find the $Q_d^2$-parameter. The results are summarized in Table 2.           
\begin{table}[h]           
\centering
\begin{tabular}{|c||c|c|c||c|c|c|}
\hline
SU(3)&\multicolumn{3}{c||}{$N_f=2$}
 &\multicolumn{3}{c|}{$N_f=3$} \\
\cline{2-7}
$\omega_d$ & $Q_d^2$ & $T_d(MeV)$ & $m(MeV)$ & $Q_d^2$ 
&$T_d(MeV)$&$m(MeV)$ \\ 
\hline 
0.7 & 0.577 & 169 & 118 & 0.400 & 140 & 98 \\
1.0 & 0.540 & 163 & 163 & 0.361 & 133 & 133 \\
1.3 & 0.523 & 160 & 208 & 0.339 & 129 & 168 \\
1.9 & 0.543 & 164 & 312 & 0.343 & 130 & 247 \\
2.5 & 0.616 & 174 & 435 & 0.395 & 140 & 350 \\
\hline
\end{tabular}
\caption{The data summary for the deconfinement phase transition 
within the SU(3)-model for $\mu=0$. We keep $\Lambda=\;$222MeV to 
estimate $T_d$. For comparison see [2] where the similar results are
discussed.} 
\end{table}

\noindent 
From Table 2 we see that indeed $T_d$=166MeV (for $N_f$=2) and 
$T_d$=140MeV (for $N_f$=3) seems to be the upper limit for this value 
since the quark mass is of the order of 350Mev or less. But in any 
case $T_d\neq T_c$ (for $T_c$ see Table 1) and this fact is more 
important. These results completely confirm the scenario where 
the Q-phase separates QGP from the usual hadronic mater: the two 
well-separated phase transitions are established with $\Delta T_{max}
=T_c-T_d\sim$30MeV.

{\it Below the case $T=0$ is investigated}. Treating this case we start
again from the two-loop approximation for $\Omega$ which for 
$T=0$ is simpler since all integrals can be calculated analytically.
Here one finds that the two-loop thermodynamical potential for cold
quark-gluon matter has the form [7]
\begin{eqnarray}
\frac{\Omega^{(2)}}{V}=
&-&\frac{NN_f\mu^4}{12\pi^2}\left[\sqrt{1-\theta^2}(1-\frac{5}{2}
\theta^2)+\frac{3}{2}\theta^4{\displaystyle\ln}\frac{1+\sqrt{1-
\theta^2}}{\theta}
\right]\nonumber\\
&+&g^2\frac{N_f(N^2-1)\mu^4}{64\pi^4}\left\{3\left[\sqrt{1-\theta^2}-
\theta^2{\displaystyle\ln}\frac{1+\sqrt{1-\theta^2}}{\theta}
\right]^2\nonumber\right.\\
&-&\left.2(1-\theta^2)^2\right\}
\end{eqnarray}
where $\theta=m/\mu$ is a new parameter and the link between 
$\mu$ and the quark density $\rho$ is given by
\begin{eqnarray}
\rho=\frac{N_f\mu^3}{3\pi^2}(1-\theta^2)^{3/2}\;.
\end{eqnarray}
At first we consider {\it the chiral phase transition} which occurs in the
quark-gluon plasma with the light (here massless) quarks.
In this limit Eq.(12) becomes very simple and, for $N=3$, 
it is reduced to [6]
\begin{eqnarray}
p=-\frac{\Omega^{(2)}}{V}=\frac{N_f\mu^4}{4\pi^2}\left(1-\frac{
g^2(Q^2)}{2\pi^2}\right)
\end{eqnarray}
where $Q=\mu/\mu_0$. The phase transition occurs 
when $p=0$ and the equation which results from this condition
has the form
\begin{eqnarray}
1=\frac{b_1}{2\;b_0^2\;{\displaystyle\ln}\Bigl[1+Q_c^{
{\displaystyle 2(\frac{b_1}{16b_0})}}\Bigr]}
\end{eqnarray}
where all notations are the same as above. This equation is 
solved numerically to find $Q_c$ and the results are summarized 
in Table 3.
\begin{table}[h]           
\centering
\begin{tabular}{|r||r|r| }
\hline
$SU(3)$ & $N_f=2$ & $N_f=3$ \\ 
\hline        
$Q_c^2$ & 0.232 & 0.195 \\
$\mu_c(MeV)$ & 700 & 618 \\
\hline
\end{tabular}                         
\caption{The data summary for the chiral phase transition
within the SU(3)-model when $T=0$. Here we choose $\mu_0=\;$1.4GeV 
to estimate the quark chemical potential $\mu_c$ in accordance with
[2,12].} 
\end{table}

\noindent
{\it The deconfinement phase transition} occurs  when the quark 
density decreases and quarks become massive: their mass is generated
on the chiral phase transition line. The critical density 
is determined in the same manner through equation (-$p=\Omega(Q_d^2)
=0$) whose explicit form is found by using Eq.(12). It is given by
\begin{eqnarray}
\frac{b_1}{b_0^2\;{\displaystyle\ln}\Bigl[1+Q_d^{
{\displaystyle 2(\frac{b_1}{16b_0})}}\Bigr]}           
=\frac{2\;\sqrt{1-\theta_d^2}(1-\frac{5}{2}\theta_d^2)
+3\;\theta_d^4\;{\displaystyle\ln}{\displaystyle\frac{1+\sqrt{1-
\theta_d^2}}{\theta_d}}}{3\left[\sqrt{1-\theta_d^2}-\theta_d^2
{\displaystyle\ln}{\displaystyle\frac{1+\sqrt{1-\theta_d^2}}{\theta_d}}
\right]^2-2(1-\theta_d^2)^2}
\end{eqnarray}
where $\theta_d=m/\mu_d$. When solving Eq.(16) one finds, at once, 
that there is an upper bound for $\theta_d$-parameter: only 
$\theta_d<$ 0.4 is acceptable to keep the right side of Eq.(16)
positive. This bound means that the quark mass generation is
suppressed when $T=0$ and we find $m_q\sim60\;-\;80$Mev only. Of 
course, this estimate depends on the scale $\mu_0$ which we fixed 
above but in any case the chiral phase transition demonstrates 
itself in the $(T=0,\mu)$-region very smoothly. Two values for the
$\theta_d$-parameter are considered below and for both cases we 
find approximately the same quark mass but completely different 
values of $\Delta\mu\;=\;\mu_c-\mu_d$ (see the results in Table 4). 
\begin{table}[h]
\centering
\begin{tabular}{|c||c|c|c||c|c|c|}
\hline
SU(3)&\multicolumn{3}{c||}{$N_f=2$}
 &\multicolumn{3}{c|}{$N_f=3$} \\
\cline{2-7}
$\theta_d$&$Q_d^2$&$\mu_d(MeV)$&$m_q(MeV)$
&$Q_d^2$&$\mu_d(MeV)$&$m_q(MeV)$ \\
\hline
1/8 & 0.148 & 538 & 67 & 0.106 & 456 & 57 \\
1/6 & 0.098 & 438 & 73 & 0.068 & 365 & 61  \\
\hline       
\end{tabular} 
\caption{The data summary for the deconfinement phase transition
within the SU(3)-model when $T=0$. Here we keep $\mu_0=\;$1.4GeV 
to estimate $\mu_d$. For comparison see [2] where the similar results 
are discussed.} 
\end{table}

\noindent           
From Table 4 one can see that the proposed scenario again 
presents two well-separated phase transitions  with a rather broad 
Q-phase but, unfortunately, the mass generation in the case $T=0$ is 
not expressed sharply.

{\it To conclude} two well-separated phase transitions have been
established within our scenario. We find that $\Delta T \sim$20MeV 
on the axis $\mu=0$ and the larger range $\Delta\mu \sim$(160-260)MeV 
on the axis $T=0$. The intermediate $(\mu,T)$-region is not 
investigated but it is obvious that the scenario should be close. 
Of course, the present scenario is only a model which should be 
improved in a selfconsistent manner but it being in agreement 
with the chiral phase transition data is a solid basis for the theory 
to come. For removing any arbitrariness it is necessary to solve 
nonperturbatively the Schwinger-Dyson equation for $m_q$ (e.g. see 
[13]) and to combine this result with the present scenario. On this 
way one determines both the dynamical quark mass $m_q(T,\mu)$ and 
the deconfinement phase transition more exactly. The fit (5) which 
we used for the running coupling constant is also should be improved. 
But this question (the same as one about the higher order corrections) 
being very important and complicated requires additional investigations. 
Here  we choose the simplest fit for $g^2(Q^2)$ to perform the lowest 
order calculations.

\newpage

\begin{center}
{\bf Acknowledgements}
\end {center}

I would like to thank Rudolf Baier and Frithjof Karsch for useful 
discussions and all the colleagues from the Department of Theoretical 
Physics of the Bielefeld University for the kind hospitality.

\begin{center}
{\bf References}
\end{center}

\renewcommand{\labelenumi}{\arabic{enumi}.)}
\begin{enumerate}

\item{ E.V.Shuryak, Phys. Lett. {\bf B 107} (1981) 103.}

\item{ O.D.Chernavskaya and E.L. Feinberg, Proc. of the 
Workshop on "Hot Dense Matter: Theory and Experiment", 
Divonne-les-Bains, June 1994,
ed. J.Letessier, H.Gutbrod and J.Rafelski (Plenum Press, 1995). }

\item{ E.S.Fradkin, Proc. Lebedev Inst. {\bf 29} (1965) 6.}

\item{A.Akhiezer and S.V.Peletminskii, Soviet Phys. JETP {\bf 11}
(1960) 1316.}

\item{ J.I.Kapusta, Nucl. Phys. {\bf B 148} (1979) 461.}

\item{ O.K.Kalashnikov, Fortschr. der Phys. {\bf 32} (1984) 525.}

\item{ T.Toimela, Intern. Journal of Theor. Phys. {\bf 24} (1985) 901.}

\item{ T.Appelquist, J.Terning and L.C.R.Wijewardhana, Preprint 
YCTP-P2-96, BUHEP-96-3, UCTP-002-96 (February 1996).}

\item{ O.K.Kalashnikov and V.V.Klimov, Phys. Lett.  {\bf B 88} 
(1979) 328.}

\item{ F. Karsch, Proc. of the Workshop "QCD-20 YEARS LATER" 
Aachen, June 1992, ed. P.M.Zerwas and H.A.Kastrup (World 
Scientific, 1993); Preprint BI-TP 95/11 (March 1995).}

\item{ A.D.Martin, R.G.Roberts  and W.J.Stirling, Phys.Rev. { \bf D 43 }
(1991) 3648; G.Altarelli, Proc. of the Workshop "QCD-20 YEARS 
LATER" Aachen, June 1992, ed. P.M.Zerwas and H.A.Kastrup (World 
Scientific, 1993).}

\item{ B.M\"uller, Nucl.Phys. {\bf A 590} (1995) 3c; T.D.Lee, Nucl.
Phys. {\bf A 590} (1995) 11c.}

\item{ O.K.Kalashnikov, Z.Phys. {\bf C 39} (1988) 427.}

\end{enumerate}

\end{document}